\documentclass[11pt,twoside]{article}

\usepackage{asp2006}
\usepackage{epsf}
\usepackage{psfig}
\usepackage{graphicx}
\usepackage{lscape}
\usepackage{amsbsy}
\usepackage{amssymb}

\begin{document}
\title{Temporal evolution of magnetic elements}

\author{R. Rezaei\altaffilmark{1}, R. Schlichenmaier\altaffilmark{1}, W. Schmidt\altaffilmark{1} and C. Beck\altaffilmark{1,2}}   

\affil{1-Kiepenheuer-Institut f\"ur Sonnenphysik, 79104 Freiburg, Germany\\
2-Instituto de Astrof\'isica de Canarias~(IAC), E 38\,205, La Laguna, Spain}    


\begin{abstract} 
We study the structure and evolution of the magnetic field of the quiet Sun
by investigating weak spectro-polarimetric signals.
 To this end, we observed a quiet region close to the disk center 
with the German VTT in Tenerife, July 07, 2006. 
We recorded 38 scans of the same area. Each scan was
eight arcsec wide and observed within about 100 seconds. 
We used POLIS to simultaneously observe Stokes profiles of the neutral 
iron lines at 630.15 and 630.25\,nm, the Stokes-$I$ profile of 
the \ion{Ca}{ii}\,H line at 396.8\,nm, and a continuum speckle channel at 500\,nm. 
We witness two examples of magnetic flux cancellation of small-scale 
opposite-polarity patches, followed by an enhanced chromospheric emission. 
In each case, the two opposite-polarity patches gradually 
became smaller and, within a few minutes, the smaller one 
completely disappeared. The larger patch also diminished significantly. 
We provide evidence for a cancellation scenario in the photosphere which leaves 
minor traces at the chromospheric level.
\end{abstract}

\section{Introduction}   
Observations indicate that most of the magnetic flux passing through the photosphere  
is concentrated in \textit{magnetic elements}, i.e., patches of high field strength that are 
embedded in relatively field-free plasma (Solanki 1993). 
The field lines of the magnetic elements are nearly vertical to the surface because of buoyancy forces. 
In the hierarchy of the 
magnetic structures, the magnetic elements take the position between dark and bright structures, 
i.e., between pores and network bright points (Zwaan 1987, Stenflo 1994). 
While the magnetic elements are not visible 
in continuum or line wing, they appear bright in the core of 
the chromospheric \ion{Ca}{ii}\,H and K lines.
The convective motion associated with
the photospheric granulation sweeps the magnetic flux toward the intergranular lanes. Due to the geometry of
these lanes, the magnetic flux is arranged there either in chains of individual flux tubes or elongated sheets. This
behavior is seen in high-resolution observations \citep{berger_etal_04, rouppe_etal_05} as well as in
numerical simulations (Steiner 2005, V\"ogler et al.\,2005).

Magnetic flux cancellations are common events in the solar atmosphere. 
There are examples in which it leads to a clear enhancement in the 
chromospheric intensity (Bellot Rubio \& Beck 2005; Beck, Bellot Rubio \& Nagata 2005). 
Magnetic reconnection at the photospheric level was also studied 
by Litvinenko (1999) and Takeuchi \& Shibata (2001). 
Rezaei et al.\,(2007b) presented Stokes-$V$ profiles of the \ion{Fe}{i}\,630\,nm line pair, 
where the two lines show opposite polarities in a single spectrum (OP profile). 
They suggested that it may be understood as a magnetic reconnection event 
at the solar photosphere with a line of arguments similar to Steiner (2000).

In this contribution, we show the temporal evolution of physical quantities before 
and after this event. In addition to the polarimetric data, we investigate 
\ion{Ca}{ii}\,H profiles which contain information about higher layers of the solar atmosphere.

\section{Observations and data reduction}
A time-series of a small quiet Sun region close to disk center 
($\cos\theta$\,=\,0.99), was observed with the VTT in Tenerife, July 07, 2006. 
The seeing was good and stable during the observation. 
The Kiepenheuer Adaptive Optics System  was used for maximum spatial 
resolution and image stability (von der L\"uhe et al.\,2003). 
For 64 minutes, we scanned an area eight arcsec wide with a scanning cadence of about 97~s. 
The scanning step size and spatial sampling along the slit were  0.5 and 0.3\,arcsec, respectively. 
The spectrograph slit was 0.48\,arcsec wide. The slit height of the blue 
(396.8\,nm) and red (630\,nm) channels  was 70 and 95\,arcsec, respectively.

Full Stokes profiles of the \ion{Fe}{I}\,630\,nm line pair  
and the Stokes-$I$ profile of the \ion{Ca}{ii}\,H line
were observed strictly simultaneously with the red  and blue 
channel of POlarimetric Littrow Spectrograph (POLIS, Schmidt et al.\,2003, Beck et al.\,2005b). 
The spectral sampling of 1.92\,pm for the blue channel  
and 1.49\,pm for the red channel leads to a velocity dispersion of 
1.45 and 0.7\,km\,s$^{-1}$ per pixel, respectively. 
The spectrograph curvature was corrected using the routine described 
in Rezaei et al.\,(2006). The spectro-polarimetric data of the red channel 
were corrected for instrumental effects and telescope 
polarization with the procedures described by Beck et al.\,(2005a,b). 
The rms noise level of the Stokes parameters in the continuum  
was  $\sigma$\,=\,6.0\,$\times 10^{-4}$\,$I_{\mathrm{c}}$. 
We normalized the \ion{Ca}{ii}\,H intensity profiles at the line wing at 396.490\,nm 
to the FTS profile (Stenflo et al.\,1984). Following Cram \& Dam\'e (1983) and 
Lites et al.\,(1993), we define a set of parameters to quantify properties of 
the \ion{Ca}{ii}\,H line profiles (Table~1). 
The normalization procedure of the calcium line profiles is 
very similar to that described by Rezaei et al.\,(2007a).

\begin{table}
\begin{center}
\caption{Definition of the characteristic parameters of 
the \ion{Ca}{ii}\,H line profile (Lites et al.\,1999). Wavelengths are in nm.}
\begin{tabular}{l r} 
\hline\hline
quantity  &  definition \\\hline
H-index  &   396.849\,$\pm$\,0.050\\ 
H$_{2\textrm{v}}$ & 396.833\,$\pm$\,0.008  \\
H$_{2\textrm{r}}$ & 396.865\,$\pm$\,0.008 \\
V/R       &   H$_{2\textrm{v}}$/H$_{2\textrm{r}}$\\
W1\,(outer wing) & 396.632\,$\pm$\,0.005  \\
W2\,(middle wing) & 396.713\,$\pm$\,0.010   \\
W3\,(inner wing) & 396.774\,$\pm$\,0.010   \\
\hline
\end{tabular}
\end{center}
\end{table}

Simultaneously to observations of spectral lines, a continuum speckle channel 
in POLIS recorded a larger field of view at 500\,nm. 
The speckle reconstruction was performed using the Kiepenheuer-Institut Speckle 
Imaging Package (Mikurda \& von der L\"uhe 2006, W\"oger 2006). 
The spatial resolution of the reconstructed image is about 0.3~arcsec (Figs.~1 and 2, column $b$). 
We used the POLIS intensity map~(Figs.~1 and 2, column $c$) 
and the reconstructed image (column $b$) to align the data.
%

\begin{figure*}[!ht]
\plotone{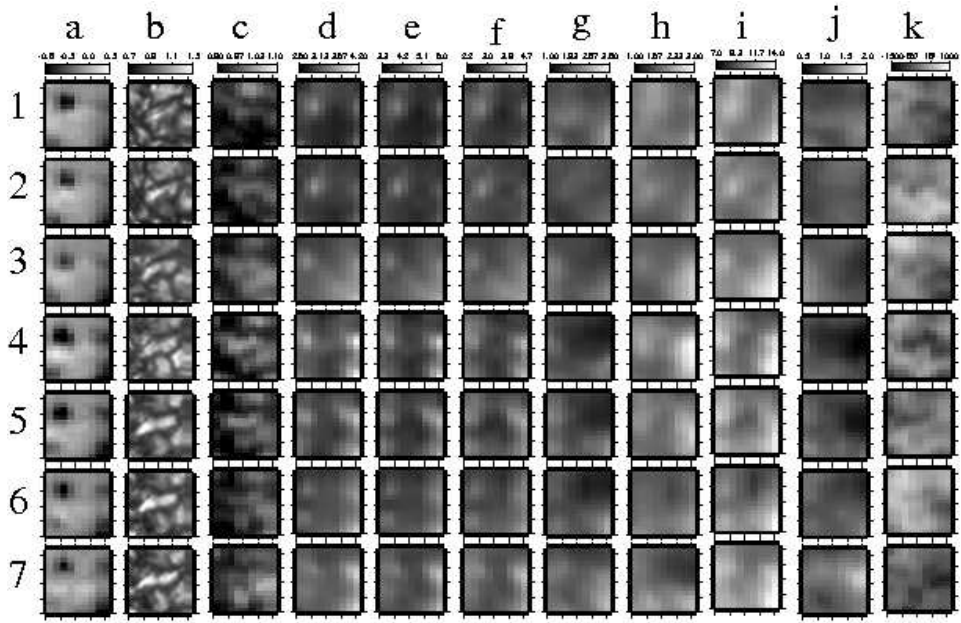}
\caption{Temporal evolution of physical parameters in a flux cancellation event. 
\emph{From left to right:} \textbf{a)} $V_{\mathrm{tot}}$, \textbf{b)} speckle reconstructed image, 
\textbf{c)} Stokes-$I$, \textbf{d)} W1, \textbf{e)} W2, \textbf{f)} W3, 
\textbf{g)} H$_{2\mathrm{v}}$, \textbf{h)} H$_{2\mathrm{r}}$, \textbf{i)} H-index, 
\textbf{j)} V/R, and \textbf{k)} \ion{Fe}{i}\,630.25\,nm velocity (see Table~1 for definitions). 
Numbers on the left show the time steps where each time step is about 97 seconds. 
The OP profile was observed in time step No. 4 ($x=1$, $y=2$). 
There are weak enhancements in the corresponding chromospheric emissions, e.g., $(4-i)$. 
The spatial extension of each map in both axes is 4.4\,arcsec.}
\label{example_1}
\end{figure*}

We used the procedure explained in the Appendix of Beck et al.\,(2007a) 
to remove effect of the differential refraction between the red and blue beams of POLIS. 
There is a time lag between co-spatial data in the two channels; 
the polarimetric data was actually recorded 10 seconds later than the calcium data 
for the case shown in Fig.~1. 
Figures~ 1 and 2 show two examples of flux cancellation event: 
one with and one without significant chromospheric brightening. 
This can be seen by comparing the maps $(4-i)$ in Fig.~1 and $(3-i)$ in Fig.~2.

We use the signed integral of the Stokes-$V$ profile, $V_{\mathrm{tot}}$, 
which traces the magnetic flux (Lites et al.\,1999; Rezaei et al.\,2007b). It enables us to 
follow weak polarimeric signals where the Stokes-$V$ amplitude is below the 3\,$\sigma$ noise level. 
The column $a$ of Figs.~1 and 2 shows the variation of $V_{\mathrm{tot}}$ for 
seven time steps around the reconnection frame (row 4). The cancellation starts as 
the smaller (white in Fig.~1) patch concentrates and lasts until it has almost disappeared. 

The V/R is a measure of the asymmetry between the violet and red emission peaks in the calcium profile. 
Signatures of bright and dark structures in V/R maps (column $j$) can be directly compared with 
H$_{2\mathrm{v}}$ and H$_{2\mathrm{r}}$ maps (columns $g$ and $h$, respectively). 
The wing intensities show the gradual variation of the intensity, 
in-between the Stokes-$I$ and the calcium core parameters. 
The photospheric velocity, column $k$, shows the patterns of up- and downflow 
structures corresponding to granules and intergranular lanes 
(e.g., compare the upper left corner of $(3-b)$ and $(3-k)$).

\begin{figure*}[!ht]
\plotone{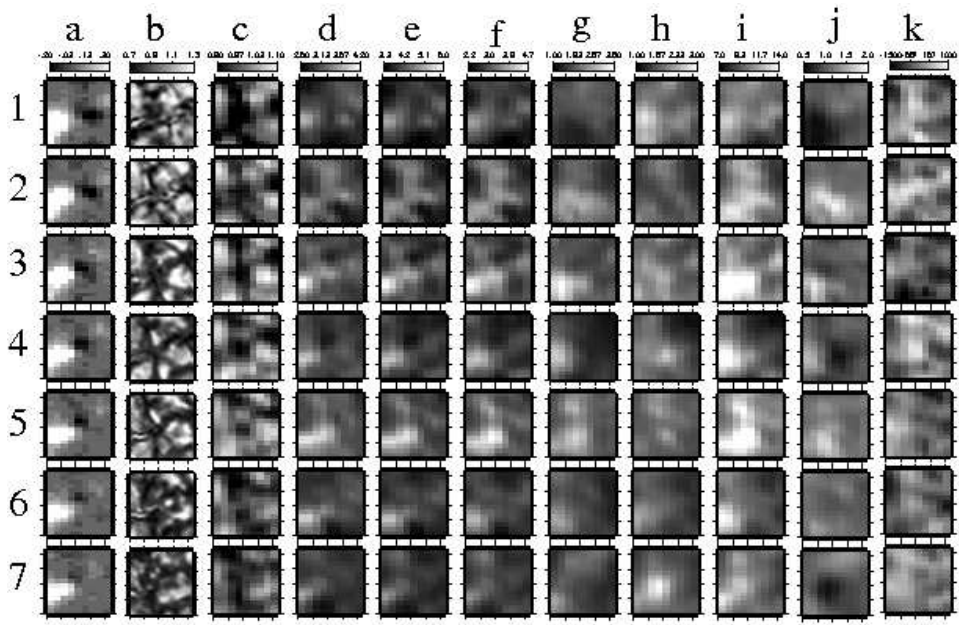}
\caption{Same as Fig.~1 but for another example of flux cancellation in our data. 
In this case, there is a significant enhancement in the chromospheric emission 
(H-index, H$_{2\mathrm{v}}$, and H$_{2\mathrm{r}}$), far stronger than the case in Fig.~1.}
\label{example_2}
\end{figure*}

\section{Discussion}

\paragraph{Similarities:} The negative polarity patch in Fig.~1 (column $a$) was a network patch, which was present during the whole observing run. In contrast, the positive patch gathered from diffuse flux, 
enhanced, and almost disappeared (the positive patch corresponds to the white color in Fig.~1, 
Rezaei et al.\,2007b). Since the magnetic flux is continuously replaced in the network, events 
like this may be common in the solar photosphere (Schrijver et al.\,1997). The strong positive 
patch in Fig.~2 (column $a$) was not a network component. However, it was much stronger than the 
negative one and persisted for the whole observing run. 
Although, the larger patches in both cases survived the cancellation event, they weakened clearly. 
In both cases, the cancellation event happened in an intergranular vertex (see panel $(4-b)$ in Figs.~1 and 2). 
Therefore, these two events have comparable configurations at the photospheric level.

\paragraph{Differences:} Comparison of the two flux cancellation events 
presented in the previous sections demonstrates essential differences in the chromospheric reaction. 
In the first case, we have mild chromospheric brightening in the cancellation site, 
e.g., row (4) in Fig.~1. In contrast, rows (3-5) of 
Fig.~2 show a brightening in the H-index on the cancellation site. Note that the colorbar has similar 
scales for the H-index. In the latter case, we observe stronger 
violet emission than red one, either by investigating columns $g$ and $h$ 
or by inspecting the asymmetry parameter, V/R in column $j$. 
In contrast, we have a stronger red emission peak in the first case (map $(4-j)$ in Fig.~1). 
Figure 3 shows the evolution of calcium profiles on and near the cancellation site. 
During the cancellation, the profiles display a H$_{2\mathrm{r}}$ peak that is much 
stronger than H$_{2\mathrm{v}}$. It is in contrast to the average quiet Sun profile where it was 
interpreted as the signature of upwards propagating acoustic waves (e.g., Beck et al.\,2007b). 
Hence, both the integrated intensity and the asymmetry parameter indicate important 
differences in the reaction of the chromospheric layers to the flux cancellation.
\begin{figure*}[!ht]
\plotone{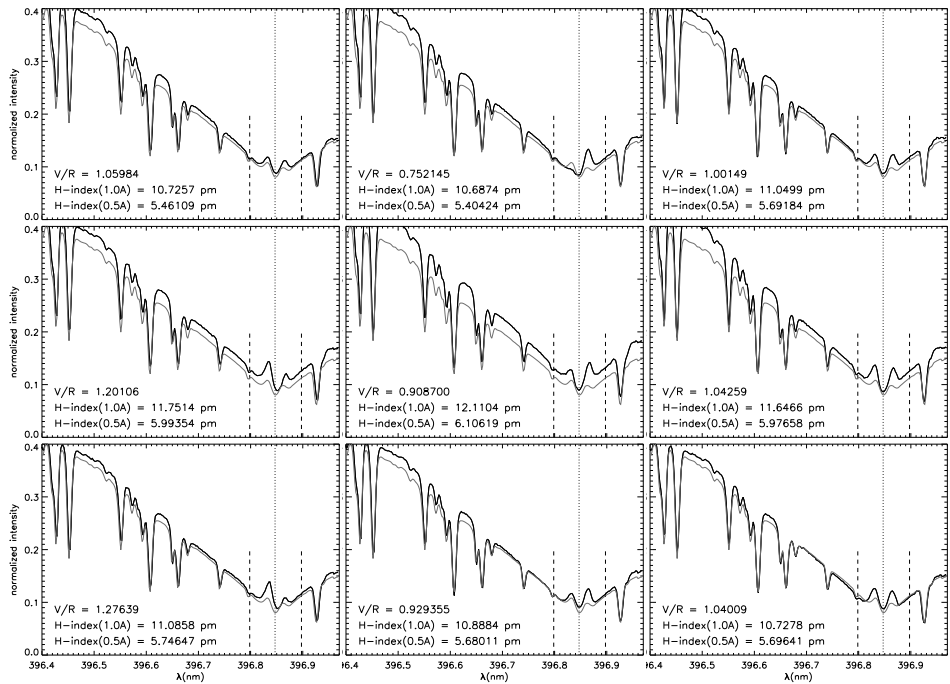}
\caption{Variation of emission peaks at three locations ([1,3] top, [0.5, 2.6] middle, [0.5, 2.2] bottom) 
in time-steps 3 (left), 4 (middle), and 5(right) for the Fig.~1 example. 
The gray  profile is the average of a few thausand profiles. The dashed lines show 
a distance of $\pm$\,0.05\,nm from the average calcium core, indicated by a vertical dotted line.}
\label{vr}
\end{figure*}

\section{Conclusions}
Time-series of co-spatial and co-temporal polarimetric data of a quiet Sun region 
revealed a variety of connections between photospheric flux cancellations 
and chromospheric enhanced emissions. 
We present two examples of magnetic flux cancellations with different 
levels of the enhanced chromospheric emission. We attribute the 
difference to the geometrical height at which the cancellation happened. 
This is in accordance with the interpretation of the 
opposite polarity profile as a signature of photospheric reconnection by Rezaei et al. (2007b).

\acknowledgements 
The POLIS instrument has been developed by the Kiepenheuer-Institut in 
cooperation with the High Altitude Observatory (Boulder, USA). 
Part of this work was supported by the Deutsche Forschungsgemeinschaft (SCHM 1168/8-1).



\end{document}